\documentstyle[12pt]{article}

\begin{document}
{\sf \begin{center} \noindent
{\Large \bf Curvature-driven diffusive and dynamo actions in vortex unstretched filaments from solar prominences data}\\[3mm]

by \\[0.3cm]

{\sl L.C. Garcia de Andrade}\\

\vspace{0.5cm} Departamento de F\'{\i}sica
Te\'orica -- IF -- Universidade do Estado do Rio de Janeiro-UERJ\\[-3mm]
Rua S\~ao Francisco Xavier, 524\\[-3mm]
Cep 20550-003, Maracan\~a, Rio de Janeiro, RJ, Brasil\\[-3mm]
Electronic mail address: garcia@dft.if.uerj.br\\[-3mm]
\vspace{2cm} {\bf Abstract}
\end{center}
\paragraph*{}
Examples of the use of Frenet curvature-driven motion of vortex
unstretched filamentary diffusion processes in plasmas are
investigated. The first example addresses the unstretched filaments
which are embedded in a steady plasma flow. The particle number
density of a weakly ionized plasma is shown to be proportional to
the total Frenet curvature. The particle number does not decays in
plasma along the filaments and is maintained against diffusion
losses. This relation is tested against the solar prominence data
for particle density of $10^{19} cm^{-3}$ and height of the order of
$10^{10}cm$ and non-thermal velocities of the order of $10^{5}
cm.s^{-1}$. Making use of a molecular diffusion constant of
$10^{6}cm^{2}s^{-1}$ one obtains a Frenet curvature of the solar
loops of the order of $10^{-8} cm^{-1}$ which coincides with the
twist (torsion) of the helical model where torsion coincides with
curvature. The second one handles a generalization of the Grunzig et
al [Phys. Plasmas 1 (1),259 (1995)] to the case a nonvanishing
plasma velocity besides the auxiliary medium velocity considered by
them. It is shown that the filaments are necessary torsion-free for
the condition that the plasma velocity has the same direction of the
auxiliary velocity flow. In this last example no dynamo action is
present and diffusion action is fully dominant despite of the
difference in velocities of plasma and auxiliary flows. This can be
better understood by the unstretched carachther of the filaments
despite of its folding or curvture. }

\newpage
\newpage
 \section{Introduction}
 Filamentary structures have been present \cite{1,2} in many physical and biological phenomena as bio-actine, solar
 and plasma filaments and dynamo filaments \cite{3}. Curvature-driven motion \cite{4} along the vortex filament or thin tubes filled with
 plasma flow is found present in solar coronal regions and prominences \cite{5,6}. Diffusion processes were throughly
 investigated by S. Molchanov \cite{7} in the seventies in Riemannian geometry and more
 recently diffusion processes have played a fundamental role in plasma physics , mainly in the case where plasma lifetime
 \cite{8} is determined by diffusion rate which varies by an order
 of  magnitude inside a hollow plasma pre-ionized by a laser beam.
 Another interesting plasma diffusion application has been recently
 given by Chen \cite{8}, by investigating the solutions of
 diffusion-convection equations which are important for magnetic
 plasma confinment. In this report one is able to investigate two new examples of the application of the vortex filaments in solar
 prominences and electrons density data is used to obtain the curvature of the solar loop which coincides with the twist
 obtained by Lopez-Fuentes et al \cite{9} of the
 magnetic twist. Another example of diffusion in plasmas is given by
 the case of auxiliary velocity in semi-ideal magnetohydrodynamics which shows that as long as the constraints on the filament
 are given by a torsion-free or planar filaments which appears frequently in tokamaks. However the new feature here is that
 contrary to the work by Grunzig et al \cite{10} is that here the plasma velocity is nonzero and diffusion is not a pure one.Therefore as a consequence of the filamentary \cite{9}
 magnetic diffusion in solar loops , flow perturbation is fully
 determined from the magnetic twist. The paper is organized as
 follows: In section 2 a brief review on holonomic Frenet frame
 is presented. In section 3 the magnetic diffusion equation is
solved in this frame for solar loops and data is seen to be
compatible with the model. Section 4 presents the generalized
Grunzig et al model for plasma filaments. Conclusions and future
prospects are presented in section 5.
\newpage
\section{Plasma vortex filaments in holonomic Frenet
frame} This section addresses a small review of the holonomic frame
\cite{11} equations that are specially useful in the investigation
of vortex filaments in MHD in homogeneous and isotropic flows as
well as isotropic magnetic diffusion. Since the magnetic filaments,
considered here, might possess Frenet torsion and curvature
\cite{11}, which completely determine topologically the filaments,
one needs some dynamical relations from vector analysis and
differential geometry of curves such as the Frenet frame
$(\vec{t},\vec{n},\vec{b})$ equations
\begin{equation}
\vec{t}'=\kappa\vec{n} \label{1}
\end{equation}
\begin{equation}
\vec{n}'=-\kappa\vec{t}+ {\tau}\vec{b} \label{2}
\end{equation}
\begin{equation}
\vec{b}'=-{\tau}\vec{n} \label{3}
\end{equation}
The holonomic dynamical relations from vector analysis and
differential geometry of curves by $(\vec{t},\vec{n},\vec{b})$
equations in terms of time
\begin{equation}
\dot{\vec{t}}=[{\kappa}'\vec{b}-{\kappa}{\tau}\vec{n}] \label{4}
\end{equation}
\begin{equation}
\dot{\vec{n}}={\kappa}\tau\vec{t} \label{5}
\end{equation}
\begin{equation}
\dot{\vec{b}}=-{\kappa}' \vec{t} \label{6}
\end{equation}
along with the flow derivative
\begin{equation}
\dot{\vec{t}}={\partial}_{t}\vec{t}+(\vec{v}.{\nabla})\vec{t}
\label{7}
\end{equation}
From these equations and the generic flow \cite{12}
\begin{equation}
\dot{\vec{X}}=v_{s}\vec{t}+v_{n}\vec{n}+v_{b}\vec{b} \label{8}
\end{equation}
one obtains
\begin{equation}
\frac{{\partial}l}{{\partial}t}=(-\kappa{v}_{n}+{v_{s}}')l\label{9}
\end{equation}
where l is given by
\begin{equation}
l:=(\vec{X}'.\vec{X}')^{\frac{1}{2}}\label{10}
\end{equation}
which shows that if $v_{s}$ is constant, which fulfills the
solenoidal incompressible flow
\begin{equation}
{\nabla}.\vec{v}=0\label{11}
\end{equation}
and $v_{n}$ vanishes, one should have an unstretched plasma vortex
filament. Is exactly this choice $\vec{v}=v_{s}\vec{t}$, of steady
flow one might choose here. Before solving the magnetic diffusive
equation in the Frenet, let us consider the form of filaments
\begin{equation}
{\partial}_{t}\vec{N}+(\vec{v}.{\nabla})\vec{N}={D}{\nabla}^{2}\vec{N}\label{12}
\end{equation}
where the parameter ${D}$ is the here considered constant diffusion
coefficient since one is considering weakly ionized plasmas, and N
is the particle number density of electrons or ionized particles in
plasmas. The solution
\begin{equation}
\vec{B}=B(s)\vec{t}\label{13}
\end{equation}
shall be considered here. This definition of magnetic filaments is
shows from the solenoidal carachter of the magnetic field
\begin{equation}
{\nabla}.\vec{B}=0\label{14}
\end{equation}
and $B_{s}$ is constant along the filaments. In the next section one
shall solve the diffusion equation in the steady case in the
non-holonomic Frenet frame as
\begin{equation}
\frac{\partial}{{\partial}n}\vec{t}={\theta}_{ns}\vec{n}+[{\Omega}_{b}+{\tau}]\vec{b}
\label{15}
\end{equation}
\begin{equation}
\frac{\partial}{{\partial}n}\vec{n}=-{\theta}_{ns}\vec{t}-
(div\vec{b})\vec{b} \label{16}
\end{equation}
\begin{equation}
\frac{\partial}{{\partial}n}\vec{b}=
-[{\Omega}_{b}+{\tau}]\vec{t}-(div{\vec{b}})\vec{n}\label{17}
\end{equation}
\begin{equation}
\frac{\partial}{{\partial}b}\vec{t}={\theta}_{bs}\vec{b}-[{\Omega}_{n}+{\tau}]\vec{n}
\label{18}
\end{equation}
\begin{equation}
\frac{\partial}{{\partial}b}\vec{n}=[{\Omega}_{n}+{\tau}]\vec{t}-[\kappa+(div\vec{n})]\vec{b}
\label{19}
\end{equation}
\begin{equation}
\frac{\partial}{{\partial}b}\vec{b}=
-{\theta}_{bs}\vec{t}-[\kappa+(div{\vec{n}})]\vec{n}\label{20}
\end{equation}
The structures ${\theta}_{ns}$, ${\Omega}_{n}$ fulfills the
following identities
\begin{equation}
{\Omega}_{n}=\vec{b}.{\partial}_{b}\vec{n}-\tau \label{21}
\end{equation}
\begin{equation}
\frac{\partial}{{\partial}b}\vec{n}={\tau}\vec{t}-[\kappa+(div\vec{n})]\vec{b}
\label{22}
\end{equation}
\section{Steady Diffusion in Solar Prominences}
To solve the diffusion equation one must express the gradient
operator along the magnetic loop in the form of nonholonomic Frenet
equations
\begin{equation}
{\nabla}=\vec{t}{\partial}_{s}+\vec{n}{\partial}_{n}+\vec{b}{\partial}_{b}\label{23}
\end{equation}
Substitution of these expressions in the magnetic diffusion equation
yields
\begin{equation}
{\nabla}^{2}={{{\nabla}^{2}}_{||}}-{\kappa}{\partial}_{n}+({\theta}_{bs}+{\theta}_{ns}){\partial}_{s}+{div\vec{b}}{\partial}_{b}
\label{24}
\end{equation}
where
${{{\nabla}^{2}}_{||}}:={{\partial}_{s}}^{2}+{{\partial}_{n}}^{2}+{{\partial}_{b}}^{2}$.
To simplify matters one considers that ${\theta}_{bs}+{\theta}_{ns}$
and $div\vec{b}$ both vanish. Substitution of these expression in
the steady plasma flow where $\frac{{\partial}N}{{\partial}t}$
vanishes, by the method of separation of variables in PDE yields
\begin{equation}
N(s,n)={\psi}(s){\phi}(n)\label{25}
\end{equation}
yields the following equations
\begin{equation}
{{\partial}_{s}}^{2}{\psi}+\frac{v_{0}}{D}{\partial}_{s}{\psi}+{k}^{2}{\psi}=0\label{26}
\end{equation}
where $v_{s}=v_{0}=constant$, and
\begin{equation}
{{\partial}_{n}}^{2}{\phi}-{{\tau}_{0}}{\partial}_{n}{\phi}-{k}^{2}{\phi}=0\label{27}
\end{equation}
where ${\tau}_{0}$ is the constant Frenet torsion and $k^{2}$ is the
separation constant. Solution of these equations yields together the
final solution
\begin{equation}
N(s,n)={k_{1}}^{2}n+\frac{D}{v_{0}}\int{{\kappa}(s)ds}\label{28}
\end{equation}
where the last term integral represents the total Frenet curvature
along the solar loop. To test this model one considers its
application to a solar prominence where non-thermal electrons
density of $10^{19} cm^{-13}$ with velocities averaged to the order
of $2.6{\times}10^{5} cm$. With these data and the height of
$10^{10} cm$ and molecular diffusivity of $10^{6} cm^{2}.s^{-1}$. By
considering helical loops where the torsion equals the Frenet
curvature and it is constant, the curvature can be computed from
these data and expression (\ref{28}) as of order $10^{-8} cm^{-1}$
which agrees with the twist or filament torsion of coronal loop
recently computed by Lopez-Fuentes et al. The vortex equation
\begin{equation}
\vec{\omega}={\nabla}{\times}\vec{v}={v_{0}}[{\kappa}\vec{b}-({\tau}+{\Omega}_{b})\vec{t}]\label{29}
\end{equation}
which along with solenoidal vorticity equation yields
\begin{equation}
{\nabla}.\vec{\omega}=v_{0}{\partial}_{s}[{\tau}+{\Omega}_{b}]=0\label{30}
\end{equation}
which yields
\begin{equation}
{\tau}=-{\Omega}_{b}+c_{1}\label{31}
\end{equation}
where $c_{1}$ is an integration constant. Now let us consider
\cite{8} the weakly ionized plasma gas where instead as in previous
solution, now one considers that These relations shall be useful now
to apply them to the diffusive plasma particles equation
\begin{equation}
D{\nabla}^{2}{N}=-Q \label{32}
\end{equation}
which considered then that in this case the source term $Q=ZN$ where
Z is the "ionization function" yields
\begin{equation}
{\nabla}^{2}{N}=-\frac{Z}{D}N \label{33}
\end{equation}
which by assuming now that N depends only on coordinate-s yields
\begin{equation}
{{\partial}_{s}}^{2}{N}+\frac{Z}{D}N=0 \label{34}
\end{equation}
and final solution
\begin{equation}
{N}=N_{0}sin[\sqrt{{Z}{D}}s] \label{35}
\end{equation}
which shows that the solution is periodic. Thus plasma is maintained
again diffusion losses along the plasma filament.
\newpage
\section{Magnetic field diffusion along plasma filaments and dynamo action}
In this section one considers the equations for the semi-ideal MHD
given by
\begin{equation}
{\partial}_{t}{\vec{B}}={\eta}{\nabla}{\times}{(\vec{u}{\times}\vec{B})}\label{36}
\end{equation}
which by considering the auxiliary flow velocity
\begin{equation}
{\vec{u}}=\vec{v}-\frac{\eta}{B^{2}}[\vec{B}{\times}({\nabla}{\times}\vec{B})]\label{37}
\end{equation}
by which considering that the grad operator now is just
$\vec{t}{\partial}_{s}$ and the $\vec{B}$ defined in the last
section one obtains
\begin{equation}
\vec{u}=\vec{v}-{\eta}{\kappa}\vec{n}\label{38}
\end{equation}
which by convenience one chooses to be $\vec{v}=v_{0}\vec{n}$ and
with this choice the equations of ideal MHD reduce to
\begin{equation}
{\kappa}{\tau}={0}\label{39}
\end{equation}
\begin{equation}
{\kappa}= v_{0}\label{40}
\end{equation}
and
\begin{equation}
{\partial}_{t}{B}_{s}=0\label{41}
\end{equation}
while the first two equations simply tells us that torsion vanishes
the last one tells us that there is no dynamo action in the plasma
despite the difference in velocities and that the diffusion is
strongly predominant.

\section{Conclusions}
A particular solution of magnetic resistivity MHD equations, is
found representing a filamentary solar loop perturbed flow which is
determined by the magnetic twist. The magnetic field is on a
magnetic resistivity setting. From the mathematical point of view,
the magnetic fields are given by holonomic filaments.Future
prospects includes the filamentary dynamo generation solutions by
the conformal geometrical technique \cite{12} of stretch,twist and
fold dynamos \cite{13} previously investigated. Speed flows involved
in plasma loops \cite{14} are extremelly bigger than the
perturbation computed here and thus it can be experimentally
disregarded, nevertheless if $v_{0}$ vanishes it is easy to see from
expression (\ref{11}) that the magnetic diffusion coefficient is a
complex number. More general framework of this work would includes
the anisotropic magnetic diffusion case. This investigation may
appear elsewhere. Curvature decay is well-within the diffusion
limits of $10^{8}yrs$ \cite{15} of the solar corona.
\section{Acknowledgements}
I appreciate financial  supports from Universidade do Estado do Rio
de Janeiro (UERJ) and CNPq (Brazilian Ministry of Science and
Technology). This paper is in memory to Professor Vladimir Tsypin ,
teacher, colleague and friend who taught us so much about
applications of Riemannian geometry in plasma physics. Fractal
geometry of filamentary dynamos can also be discussed in near future
much in the same way was done earlier by Vainshtein et al \cite{16}.

\end{document}